\documentclass[twocolumn,showpacs,prb,amsmath,amssymb]{revtex4}
\usepackage{color}
\usepackage{verbatim}
\usepackage{hyperref}
\usepackage{epsf}
\usepackage{graphics}
\usepackage{graphicx}
\usepackage{natbib}
\usepackage{dcolumn}
\usepackage{bm}

\begin{document}
\title{Meta-nematic transitions in a bilayer system: Application to the bilayer ruthenate}
\author{Christoph Puetter}
\affiliation{Department of Physics, University of Toronto, Toronto, 
Ontario M5S 1A7 Canada}
\author{Hyeonjin Doh}
\email{hdoh@physics.utoronto.ca}
\affiliation{Department of Physics, University of Toronto, Toronto, 
Ontario M5S 1A7 Canada}
\author{Hae-Young Kee}
\email{hykee@physics.utoronto.ca}
\affiliation{Department of Physics, University of Toronto, Toronto, 
Ontario M5S 1A7 Canada}
\date{\today}
\begin{abstract}
It was suggested that the two consecutive metamagnetic transitions 
and the large residual resistivity discovered in Sr$_3$Ru$_2$O$_7$ 
can be understood 
via the nematic order and its domains in a single layer system.
However, a recently reported anisotropy between two longitudinal
resistivities induced by tilting the magnetic field away from the $c$ axis
cannot be explained within the single layer nematic picture.
To fill the gap in our understanding within the nematic order scenario,
we investigate the effects of bilayer coupling and in-plane magnetic field
on the electronic nematic phases in a bilayer system.
We propose that the in-plane magnetic field in the bilayer
system modifies the energetics of the domain formation, since
it breaks the degeneracy of two different nematic orientations.
Thus the system reveals a pure nematic phase with
a resistivity anisotropy in the presence of an in-plane magnetic field.
In addition to the nematic phase,
% with distorted Fermi surfaces, 
the bilayer coupling opens a different route to a hidden nematic phase 
that preserves the $x$-$y$ symmetry of the Fermi surfaces.
\end{abstract}

\maketitle

\section{Introduction}
Electronic liquid crystal phases
have been widely discussed in the context of doped Mott insulators, 
\cite{Kivelson98Nature,Kivelson03rmp} and 
have attracted much attention with the discovery of anisotropic 
quantum Hall phases in GaAs 
heterostructures in large magnetic fields.\cite{Lilly99prl,Du99ssc,Cooper04prl}
In particular, the anisotropy in the longitudinal magnetoresistivities 
in the bilayer ruthenate Sr$_{3}$Ru$_{2}$O$_{7}$ strongly suggests the existence of 
an anisotropic metallic phase dubbed an electronic nematic phase.\cite{Borzi07Science}

In experiments on ultrapure Sr$_{3}$Ru$_{2}$O$_{7}$ 
in magnetic fields along the $c$ axis, 
an unusual phase characterized by a pronounced residual resistivity
emerges in the vicinity of a putative quantum critical point.
Furthermore, this phase is bounded by two consecutive meta-magnetic transitions.
\cite{Grigera01Science,Perry01prl,Gegenwart01prl,Perry05jps,Grigera04Science}
Upon tilting the field slightly towards one of the 
in-plane crystal axes, a magnetoresistive anisotropy appears,
where the pronounced 
resistive anomaly parallel to the in-plane field direction remains unchanged,
but disappears in the perpendicular direction.

It was proposed that the two consecutive meta-magnetic transitions 
occur due to the formation of nematic order.\cite{KeeHY05prb} 
As a consequence of the nematic order, one expects to find
an anisotropy in the longitudinal resistivities in the pure nematic phase
due to the Fermi surface distortion, but this 
has not been observed in Sr$_3$Ru$_2$O$_7$ in magnetic fields
along the $c$ axis. 
Instead, the pronounced 
resistivity shows up in the proposed nematic region bounded by
two metamagnetic transitions.
The large resistivity was explained by scattering due to
domains of two degenerate nematic orientations. \cite{DohH07prl}
However, when the magnetic field is tilted away from the $c$ 
axis, the transport 
anisotropy is discovered.
This cannot be understood within the previous nematic order proposal,
since the Zeeman coupling and the energetics of the domains
are independent of the magnetic field direction in the single layer system.

In this paper, we attempt to understand the recently reported magnetoresistive anisotropy 
in the presence of an in-plane magnetic field. 
We show that the magnetoresistive anisotropy in the presence of an in-plane magnetic field
can be understood within the nematic order picture
when one takes into account the bilayer coupling.
Note that the in-plane magnetic field in the bilayer lattice causes 
a relative momentum mismatch
between the layers. When the in-plane field is along one of the crystal axes,
it breaks the degeneracy of two different nematic orientations.
Consequently, domains with different nematic orientations are no longer 
energetically favorable,
and the system exhibits a pure nematic phase with
a Fermi surface elongation. Thus, the anisotropy in transport is 
recovered in the presence of an in-plane magnetic field.

The paper is organized as follows. 
We introduce the bilayer model in Sec. II.
In Sec. III, we identify the distinctly different nematic phases and 
present the phase diagram as a function of the bilayer coupling and chemical potential.
We also discuss a hidden nematic phase which is absent in the single layer system.
An in-plane magnetic field is incorporated in Sec. IV, where we 
study the phase diagram under the in-plane magnetic field and 
the signatures of meta-nematic transitions
in the longitudinal conductivity and the magnetic susceptibility.
We close with a discussion and a summary of our findings in Sec. V.

\section{Bilayer model}

In the electronic nematic phase, electron momenta prefer to be aligned along a
certain direction, typically along one of the crystal axes, thus breaking a point-group
symmetry of the underlying lattice. 
A number of models have been employed to study the formation 
of electronic nematic order. 
The approach adopted in the present work 
is based on the idea of a broken symmetry state of an isotropic liquid.
\cite{Yamase00jpsj,Halboth00prl,Oganesyan01prb,KeeHY03prb,Khavkine04prb,Nilsson05prb,Quintanilla06prb,Yamase05prb,Lawler06prb,MLawler06prb,Metzner03prl,DellAnna06prb,Edegger06prb,Wu07prb}
Here, the formation of the nematic phase is due to a spontaneous 
Fermi surface distortion often referred to as a Pomeranchuk instability. \cite{Pomeranchuk58jetp} 
It was found, however, 
that the divergence of the nematic susceptibility, which defines the Pomeranchuk 
instability, is preempted by a first order transition, and that 
the formation of an electronic nematic phase on a lattice is 
intimately connected to the van Hove singularities 
in the density of states (DOS).\cite{KeeHY03prb,Khavkine04prb}

The effective nematic interaction successfully describes several novel phenomena 
observed in Sr$_{3}$Ru$_{2}$O$_{7}$.\cite{KeeHY05prb,DohH07prl}
We generalize the quadrupole density interaction for a bilayer square lattice as follows,
\begin{eqnarray}
  \label{eq:NematicInteraction}
  H_{\text{int}} = && \sum_{\lambda, {\bf q}, \sigma} \, F_{2}({\bf q}) \, 
  \text{tr}[Q^{\dagger \, (\lambda)}_{\sigma}({\bf q}) \, Q^{(\lambda)}_{\sigma}({\bf q})] \nonumber \\
  &+& \sum_{\lambda, {\bf q}, \sigma} \, G_{2}({\bf q}) \, 
  \text{tr}[Q^{\dagger \, (\lambda)}_{\sigma}({\bf q}) \, Q^{(-\lambda)}_{\sigma}({\bf q})], 
\end{eqnarray}
where the symmetric and traceless tensors $Q^{(\lambda)}_{\sigma}$ are given by
\begin{eqnarray}
  &&Q^{(\lambda)}_{\sigma}({\bf q}) = 
%\frac{1}{L} 
\sum_{\lambda, {\bf k}, \sigma} 
c^{\dagger \, (\lambda)}_{{\bf k} + {\bf q}/2, \sigma}  \\
  &&\! \! \times
  \left[ \!
  \begin{array}{cc}
    \! \text{cos}(k_{x}) - \text{cos}(k_{y}) & \text{sin}(k_{x}) \, 
    \text{sin}(k_{y}) \! \\
    \! \text{sin}(k_{x}) \, \text{sin}(k_{y}) & \text{cos}(k_{y}) 
    - \text{cos}(k_{x}) \!
  \end{array}
  \! \right] 
  \! \! c^{(\lambda)}_{{\bf k} - {\bf q}/2, \sigma}. \nonumber
\end{eqnarray}
Here, $\lambda = \pm 1$ denotes the layer index,  $\sigma = \pm 1$
the spin degree of freedom, and  $c^{\dagger}$, $c$ the
electronic creation and annihilation operators.
The functions $F_{2}({\bf q})$ and $G_{2}({\bf q})$ denote
intraplane and  interplane quadrupolar density interactions, respectively.
%The linear system size is given by $L$, which we set to unity in the following.
The order parameter is defined through the tensors 
$\langle Q^{(\lambda)}_{\sigma} \rangle$
in analogy to its counterpart in classical liquid crystal theory.
In conjunction with the tight-binding model on a square lattice, 
the intralayer interaction term in Eq. (\ref{eq:NematicInteraction})
describes the first order transition between isotropic and nematic states
within a mean-field theory. 
\cite{KeeHY03prb, Khavkine04prb}

In general, there are two distinct nematic orders in a single layer square lattice system.
The preferred direction of the electron momenta can be aligned 
either parallel or diagonal to the crystal axes.
Previous studies of monolayer systems have shown, however,
that diagonal order is generally suppressed for the model given by 
Eq. (\ref{eq:NematicInteraction}).
\cite{DohH06prb,Kao05prb,Kao07prb}
We, therefore, concentrate on the nematic phase 
parallel to the in-plane crystal axes.
Assuming an attractive interlayer interaction potential
$F_{2}({\bf q}) = - F_{2} \, \delta_{{\bf q}, {\bf 0}}$
($F_{2} > 0$)
and a generally attractive intralayer interaction 
$G_{2}({\bf q}) = - G_{2} \, \delta_{{\bf q}, {\bf 0}}$,
the components of the parallel order parameter are defined by
\begin{equation}
  \label{eq:OPDefinition}
  \Delta^{(\lambda)}_{\sigma} = F_{2} \, \langle Q^{(\lambda)}_{\sigma, xx}
  ({\bf q} = 0) \rangle
  = - F_{2} \, \langle Q^{(\lambda)}_{\sigma, yy}({\bf q} = 0) \rangle,
\end{equation}
%\begin{equation}
%  \label{eq:OPDefinition}
%  \Delta = \frac{F_{2}}{L} \langle Q_{xx}({\bf q} = 0) \rangle
%  = - \frac{F_{2}}{L} \langle Q_{yy}({\bf q} = 0) \rangle,
%\end{equation}
where a positive (negative) value signals,
that electron momenta are 
preferentially aligned along the $y$ ($x$) axis.
Note that the order parameter is defined layerwise ($\lambda = \pm 1$).
Within the mean-field (MF) approximation, the Hamiltonian then becomes
\begin{eqnarray}
  H_{\text{MF}} &=& 
  \sum_{\lambda, {\bf k}, \sigma} \, 
  \varepsilon^{(\lambda)}_{{\bf k}, \sigma} \: 
  c_{{\bf k}, \sigma}^{(\lambda) \, \dagger} 
  c_{{\bf k}, \sigma}^{(\lambda)}
  + \sum_{{\bf k}, \sigma}
  t_{\perp} \, \big(c_{{\bf k}, \sigma}^{(1) \, \dagger} 
  c_{{\bf k} + {\bf p}, \sigma}^{(-1)} + \text{H.c.} \big)\nonumber \\
  &&+ \sum_{\lambda, \sigma} \, 
  \bigg\{ \frac{(\Delta^{(\lambda)}_{\sigma})^{2}}{2 \, F_{2}} \, 
  + \, G_{2} \, \frac{\Delta^{(\lambda)}_{\sigma} \, 
    \Delta^{(- \lambda)}_{\sigma}}
  {4 \, F_{2}^{2}} \bigg\},
\end{eqnarray}
%\begin{eqnarray}
%  H_{\text{MF}} &=&  \frac{1}{L^{2}} 
%  \sum_{\lambda, {\bf k}, \sigma} \, 
%  \varepsilon^{(\lambda)}_{{\bf k}, \sigma} \: 
%  c_{{\bf k}, \sigma}^{(\lambda) \, \dagger} 
%  c_{{\bf k}, \sigma}^{(\lambda)} \\
%  &&+ \frac{1}{L^{2}} \sum_{\lambda, {\bf k}, \sigma}
%  t_{\perp} \, c_{{\bf k}, \sigma}^{(\lambda) \, \dagger} 
%  c_{{\bf k} + {\bf Q}, \sigma}^{(-\lambda)} \nonumber \\
%  &&+ \sum_{\lambda, \sigma} \, 
%  \Big\{ \frac{(\Delta^{(\lambda)}_{\sigma})^{2}}{2 \, F_{2}} \, 
%  + \, G_{2} \, \frac{\Delta^{(\lambda)}_{\sigma} \, \Delta^{(- \lambda)}_{\sigma}}
%  {4 \, (F_{2})^{2}} \Big\}, 
%\end{eqnarray}x
where $t_{\perp}$ is the interlayer hopping amplitude.

To consider the effect of an in-plane magnetic field, 
we incorporate  the field via Peierls substitution 
and Zeeman coupling.
Assuming that ${\bf B} = B_{x} \, \hat{{\bf x}}$, the magnetic flux 
in the $\text{x}$-direction through a single plaquette
is given by $\phi_{x} = B_{x} a^{2}$, 
as the layer separation in Sr$_{3}$Ru$_{2}$O$_{7}$ is of about 
the same order as the planar lattice parameter $a$.
Normalized by the flux quantum $\phi_{0} = h c / e$, the flux causes the 
following relative momentum mismatch between the layers:
\begin{equation}
  {\bf p} = \frac{2 \pi}{a} \, \frac{\phi_{x}}{\phi_{0}} \hat{{\bf y}}.
\end{equation}
The electronic dispersions for each layer
in the presence of an in-plane magnetic field are given by
\begin{eqnarray}
  \label{eq:Pre_HybridDispersions}
  \varepsilon^{(\lambda)}_{{\bf k}, \sigma} 
  &=& - 2 t \, 
  \big[ 
  \text{cos}\big(k_{x} \big) 
  + \text{cos}\big(k_{y} \big) 
  \big] \nonumber \\
  && - \bigg(
  \Delta^{(\lambda)}_{\sigma} + \frac{G_{2}}{2 F_{2}} \Delta^{(-\lambda)}_{\sigma} 
  \bigg) 
  \Big[ 
  \text{cos}\big(k_{x} \big) 
  - \text{cos}\big(k_{y} \big) 
  \Big] \nonumber \\
  && - \mu - \gamma \, \pi \, \frac{\phi_{x}}{\phi_{0}} \, \sigma,
\end{eqnarray}
where ${\bf k} \rightarrow {\bf k} + {\bf p}$ for the lower layer ($\lambda = -1$). 
In Eq. (\ref{eq:Pre_HybridDispersions}), we introduced 
the chemical potential $\mu$ and an effective magnetic moment $\gamma$
($=g \mu_B \frac{1}{a^2} \frac{\hbar}{e}$),
while $\sigma = \pm 1$ stands for spin-up and spin-down.
Taking into account the bilayer coupling $t_{\perp}$, the 
hybridized energy bands read
\begin{equation}
  \label{eq:HybridDispersions}
    E^{(\pm)}_{{\bf k}, {\bf p}, \sigma} 
    = \frac{\varepsilon^{(1)}_{{\bf k}, \sigma} 
    + \varepsilon^{(-1)}_{{\bf k}+{\bf p}, \sigma}}{2}
    \pm \sqrt{\frac{\big( \varepsilon^{(1)}_{{\bf k}, \sigma} 
      - \varepsilon^{(-1)}_{{\bf k}+{\bf p}, \sigma}\big)^{2}}{4} 
      + t_{\perp}^{2}},
\end{equation}
where the order parameter components
satisfy the set of mean-field equations ($\lambda = \pm 1$)
\begin{equation}
  \label{eq:Self-ConEq}
  \Delta^{(\lambda)}_{\sigma} = F_{2} \, \sum_{{\bf k}, \sigma} 
  \Big[ 
  \text{cos}\big( k_{x} \big) 
  - \text{cos}\big( k_{y} \big) 
  \Big] \langle c_{{\bf k}, \sigma}^{(\lambda) \, \dagger} \
  c_{{\bf k}, \sigma}^{(\lambda)} \rangle.
\end{equation}
%\begin{equation}
%  \label{eq:Self-ConEq}
%  \Delta^{(\lambda)}_{\sigma} = \frac{1}{L^{2}} \sum_{{\bf k}, \sigma} 
%  \Big( 
%  \text{cos}\big( k_{x} \big) 
%  - \text{cos}\big( k_{y} \big) 
%  \Big) \langle c_{{\bf k}, \sigma}^{(\lambda) \, \dagger} \
%  c_{{\bf k}, \sigma}^{(\lambda)} \rangle.
%\end{equation}
The expectation values depend in a non-linear fashion 
on all $\Delta^{(\lambda)}_{\sigma}$.
%Note that the components $\Delta^{(\lambda)}_{\sigma}$
%characterize the topologies of the \emph{unhybridized} Fermi surfaces.
Both dispersions of Eq. (\ref{eq:HybridDispersions})
are mapped onto each other under a particle-hole transformation.

In the following, we calculate the free energy using 
an adaptive integration scheme
and solve the set of mean-field equations self-consistently.
We first present the effects of the bilayer couplings $t_{\perp}$ and $G_2$
on nematic order in the absence of a magnetic field 
($\phi_{x} = 0$).
Based on the zero-field results, we then deduce the effect of an
in-plane magnetic field originating from the bilayer coupling $t_{\perp}$.
Finally, we include the Zeeman term and investigate
the signatures of nematic order in magnetization and transport.

\section{Nematic phases in a bilayer system  
\label{section:NematicPhasesInABilayer}}

\begin{figure}[tt]
  \includegraphics*[angle=0, width=1.0\linewidth, clip]{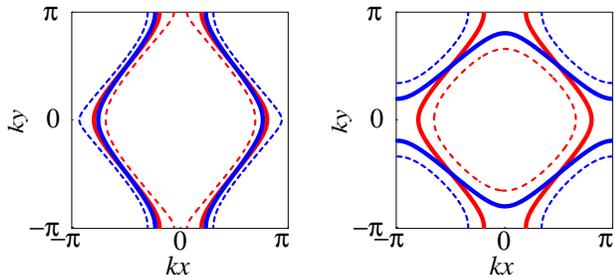}
  \caption{(Color online) Fermi surface shapes of the two nematic phases 
    in the bilayer system. The solid lines represent the prehybridized 
    and the dotted lines the hybridized Fermi surfaces. The left panel shows 
    the Fermi surface configuration in the parallel nematic phase,
    where both prehybridized Fermi surfaces are elongated
    parallel to each other. The right panel shows the hidden nematic phase,
    where both prehybridized Fermi surfaces 
    are elongated perpendicular to each other,
    while the hybridized Fermi surfaces recover the fourfold symmetry.
    \label{fig:twoFS}}
\end{figure}

Let us first identify possible nematic phases
in the absence of external fields ($\phi_{x} = 0$).
The spin index is redundant and the state of the system is specified 
by a layer-dependent order parameter $\Delta^{(\lambda)}$ ($\lambda = \pm 1$).
As none of the crystal directions is distinguished, 
both components of the order parameter 
have the same magnitude, $|\Delta^{(1)}| = |\Delta^{(-1)}|$.
This implies that, besides the isotropic phase, 
where $\Delta^{(1)} = \Delta^{(-1)} = 0$, 
only two\cite{footnote} distinct nematic phases can occur as shown in Fig. 1:
a parallel nematic phase, where $\Delta^{(1)} = \Delta^{(-1)}$
and a "hidden'' nematic phase, where $\Delta^{(1)} = -\Delta^{(-1)}$.
In the hidden nematic phase, the prehybridized Fermi surface of one layer 
$\varepsilon^{(\lambda)}_{\bf k}$ is elongated along the $x$ ($y$) direction,
while the prehybridized Fermi surface of the other layer  $\varepsilon^{(-\lambda)}_{\bf k}$ 
is stretched along the $y$ ($x$) axis.
However, taking into account the bilayer coupling, each of the hybridized
energy bands shown as the thick lines in Fig. \ref{fig:twoFS}
preserves the \mbox{$x$-$y$ symmetry}, but breaks the relative symmetry
between the layers, $(\Delta^{(1)}-\Delta^{(-1)}) \neq 0$.
A similar phase called the $\alpha$ phase, 
where up-spin and down-spin Fermi surfaces are
elongated perpendicular to each other, was reported.\cite{Wu07prb}

\begin{figure}[tt]
  \includegraphics*[angle=0, width=1.0\linewidth, clip, bb=0 0 15cm 16cm]{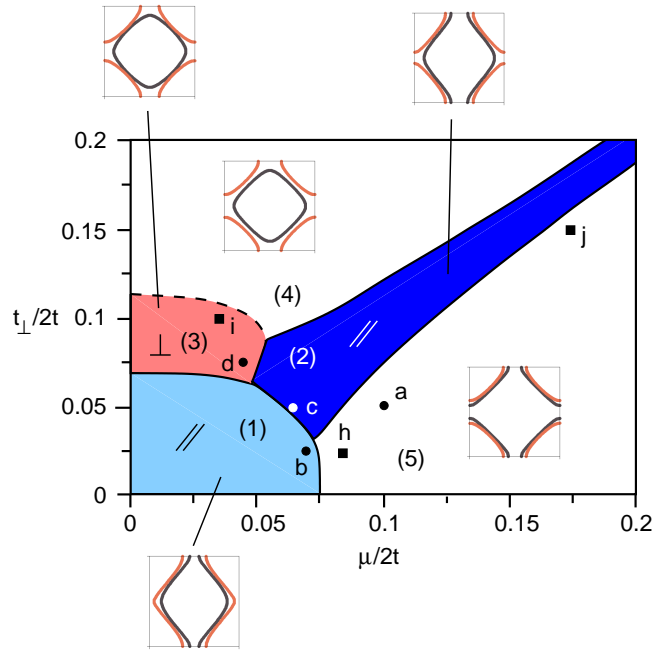}
  \caption{(Color online) Phase diagram of the bilayer system
    as a function of inter-layer hopping $t_{\perp}/2t$ and 
    chemical potential $\mu/2t$ for $F_{2}/2t = 0.8$, $G_{2}/F_{2} = 0.1$
    and $\phi_{x} / \phi_{0} = 0$. 
    The parallel nematic phase is labeled by $//$ [regions (1) and (2)] 
    and the hidden nematic phase by $\perp$ [region (3)].
    The white areas denoted by (4) and (5) are regions of isotropic phase.  
    All solid lines are first order boundaries, 
    while the dashed line represents a second order line.
    The insets show representative Fermi surface topologies
    of the hybridized energy bands  for different parts of the phase diagram.
    Note the different Fermi surface configurations 
    for parallel order in regions (1) and (2).
    The phase diagram is symmetric in $\mu$ (due to particle-hole
    symmetry) and in $t_{\perp}$.
     The DOS at various points (a)-(d) is shown in Fig. \ref{fig:DOS_1}, and 
     the conductivity and magnetization at points (h)-(j) are shown 
     in Fig. \ref{fig:ConMagSusc_hConMagSusc_iConMagSusc_j}.
    \label{fig:PhaseDiagram2D_1h}}
\end{figure}

We study the phase diagram as a function of 
bilayer coupling $t_{\perp}$ and chemical potential $\mu$.
The phase diagram shown in Fig. \ref{fig:PhaseDiagram2D_1h}
is obtained for $F_{2}/2t = 0.8$ and $G_{2}/F_{2} = 0.1$.  
%It is symmetric in $\mu$ and $t_{\perp}$,
It shows that the parallel nematic phase, labeled by (1) and (2) in Fig. \ref{fig:PhaseDiagram2D_1h}, 
is favored along the diagonal region ($\mu \sim t_{\perp}$) of the phase diagram.
The hidden nematic phase (3) on the other hand, emerges as an intermediate phase
at intermediate values for $t_{\perp}$ and is separated 
by a second order phase boundary from the neighboring isotropic regime (4).
All the other phase boundaries in Fig. \ref{fig:PhaseDiagram2D_1h}
are of first order, and involve a sudden change in the magnitude and/or orientation 
of the Fermi surface distortion dubbed nematicity.
Note that, while the nematicity suddenly changes,
there is no further symmetry breaking associated with the transition
from one nematic phase to another. 
We call such a transition a ``meta-nematic" transition in analogy to a meta-magnetic
transition, where the magnetization jumps  without any 
further symmetry breaking.
The insets in Fig.  \ref{fig:PhaseDiagram2D_1h} display the different  
Fermi surface topologies associated with the nematic phases in the bilayer system. 

\begin{figure*}[tt]
\includegraphics*[angle=0, width=0.9\linewidth, clip]{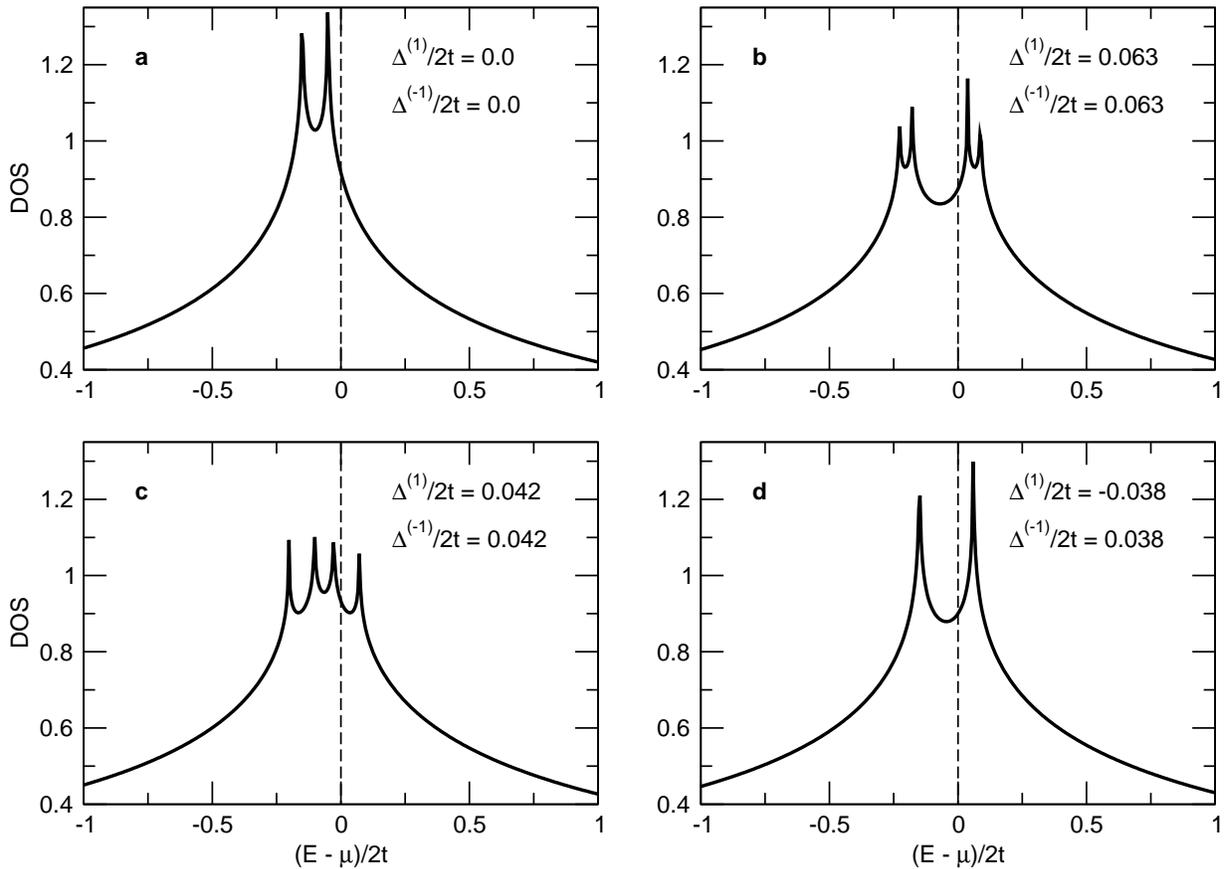}
  \caption{DOS for 
    the points (a) - (d) as marked in Fig. \ref{fig:PhaseDiagram2D_1h} (DOS in units of 2t/E).
    The parameters are fixed at 
    $F_{2}/2t = 0.8$, $G_{2}/F_{2} = 0.1$ and $t_{\perp}/2t = 0.1$.
    The order parameters have been determined self-consistently 
    through Eq. (\ref{eq:Self-ConEq}).
    The formation of parallel nematic order splits 
    the van Hove singularities into four peaks as shown in (b) and (c), while   
    the hidden nematic order shifts the two peaks further apart from each other as shown in (d).
    \label{fig:DOS_1}}
\end{figure*}

To gain a better understanding of the different types of nematic order in the  phase diagram,
we study the behavior of the DOS, since it was shown that  nematic order 
develops in order to avoid a van Hove singularity.\cite{KeeHY03prb}
In Fig. \ref{fig:DOS_1}, we present the DOS 
for the points (a)-(d) as marked in Fig. \ref{fig:PhaseDiagram2D_1h}.
%We choose the points (a)-(d) close to the phase boundary in order to
%show not only the effect of the nematic order 
%in a given phase but also the possibility of another transition to a nearby phase.
Originating from the underlying tight-binding dispersion,
the bilayer DOS exhibits two peaks in the absence of nematic order, 
which are separated by $2 t_{\perp}$ as shown  in Fig. \ref{fig:DOS_1} (a).

As we discussed above, nematic order prevents 
the Fermi level from lying at the van Hove singularity. 
However, there is more than one channel to avoid a van Hove
singularity in the bilayer system.
Each of the original peaks can split into two new singularities, which leads to
the parallel nematic phase and is similar to the single layer case.
In total, this gives rise to four singularities
as shown in Fig. \ref{fig:DOS_1} (b) and (c), where
the separation between two new peaks is given by
$4 \, (1 + \frac{G_2}{2F_2}) \, |\Delta^{(\lambda)}|$.
In the hidden nematic phase, in contrast, 
the two original peaks shift further away from each other such that the 
mutual separation becomes greater than the bare hybridization $2 t_{\perp}$ as shown
in Fig. \ref{fig:DOS_1} (d). 
%This is energetically more favorable than a splitting
%of singularities when $t_{\perp} > \mu$.
The peak separation then becomes 
 $2 \, \sqrt{4 \, (1-\frac{G_2}{2 F_2})^{2} \, |\Delta^{(\lambda)}|^{2} + (t_{\perp})^2 }$,
which is always greater than $2 t_{\perp}$.
This channel, leading to the hidden nematic phase, is absent in a monolayer system.

\begin{figure}[tt]
  \includegraphics*[angle=0, width=1.1\linewidth, clip, bb=0 0 16cm 16cm]{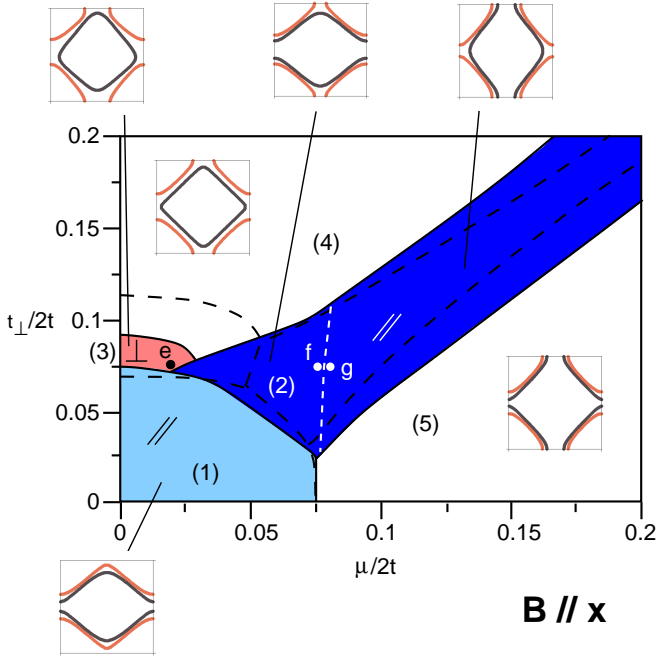}
  \caption{(Color online) Phase diagram in the presence of a 
    finite magnetic field ($\pi \, \phi_{x}/\phi_{0} = 0.2$) 
    as a function of the inter-layer hopping $t_{\perp}$ and chemical potential $\mu$ 
    for $F_{2}/2t = 0.8$ and $G_{2}/F_{2} = 0.1$.    
    No coupling of the magnetic field to the spins is assumed.
    All phase boundaries (solid lines) are of first order, as expected.  
    The dashed lines indicate the phase boundaries in 
    the absence of a field (cf. Fig. \ref{fig:PhaseDiagram2D_1h}). 
    The magnetic field determines the direction of Fermi surface elongation.
    In the parallel phase (2), a rotation of the Fermi surfaces occurs across the white dashed line. 
    The insets show the typical Fermi surface shapes encountered in 
    various regions of the phase diagram. 
    The DOS at the points (e)-(g) are plotted in Fig. \ref{fig:DOSAndFS_3a}.
   % Points (h)-(j) are the starting points at zero magnetic field for the plots
   % in Fig. \ref{fig:ConMagSusc_hConMagSusc_iConMagSusc_j}.
    \label{fig:PhaseDiagram2D_3b}}
\end{figure}

%The hybridized energy bands, which obey particle-hole symmetry, 
%take on the simplified form 
%\begin{eqnarray}
%  \label{eq:HybridDispersionsNoMagField}
%  && E^{(\pm)}_{{\bf k}} 
%  = - 2 t \, f^{+}_{{\bf k}} 
%  - \frac{1}{2} \big(1 + \frac{G_{2}}{2 F_{2}}\big) \, 
%  \big(\Delta^{(1)} + \Delta^{(-1)} \big)
%  f^{-}_{{\bf k}} \nonumber \\
%  && \pm \sqrt{ \frac{1}{4} \big( 1 - \frac{G_2}{2 F_2}\big)^{2} 
%    \big(\Delta^{(1)} - \Delta^{(-1)}\big)^{2} 
%  (f^{-}_{{\bf k}})^{2} + t_{\perp}^{2} },
%\end{eqnarray}
%where 
%$f^{\pm}_{{\bf k}} = \text{cos}\big( k_{x} \big) 
%\pm \text{cos}\big( k_{y} \big)$
%has been introduced for brevity.

%We also study the effect of the inter-layer nematic interaction $G_2$ by varing $G_2/F_2$.
%The hybridization $t_{\perp}$ and the attractive inter-layer nematic interaction 
%have converse effects, which give rise to competing nematic and isotropic phases. 
%A finite $t_{\perp}$
%favors the isotropic phase like the intra-layer hopping $t$, 
%while $G_{2}$ prefers parallel order over isotropic or hidden nematic
%phase, which is expected.

\section{Effect of in-plane magnetic field}
\label{section:EffectOfIn-planeMagneticField}
As we discussed above, the two consecutive metamagnetic transitions and 
the large residual resistivity bounded by the metamagnetic transitions
can be understood within the nematic order proposal for a single layer
square lattice.
However, the recently discovered anisotropy between two longitudinal resistivities
induced by tilting the magnetic field away from the c-axis cannot be explained within
the nematic order scenario in a single layer, 
since the Zeeman coupling and the energetics of possible domains
in a single layer system
are independent of the magnetic field direction.

However, note that Ru in Sr$_3$Ru$_2$O$_7$ forms a bilayer layer square lattice, and 
the in-plane magnetic field has a dramatic effect in the bilayer lattice.
For example, a field $B_{x} \, \hat{{\bf x}}$ 
%\begin{equation}
%{\bf A} = B_x z \hat{{\bf y}}
%\end{equation}
leads to  a $k_y$ mismatch between the upper and lower layer, where the 
momentum difference ${\bf p} = 2 \pi/a \, (\phi_{x}/\phi_{0}) \, \hat{{\bf y}}$, while
$k_x$ remains unchanged.
The $x$-$y$ symmetry breaking field lifts the degeneracy between the two
nematic orientations, and thus the phase separation with domains
is no longer energetically favorable. 
Hence the system recovers a pure nematic phase with an anisotropy in the longitudinal transport. 
Before we present  signals of the anisotropy in various  quantities,
let us first study how the in-plane magnetic field affects the phase diagram
of Fig. \ref{fig:PhaseDiagram2D_1h}.
The phase diagram in the presence of an in-plane magnetic field
($\pi \, \phi_{x}/\phi_{0} = 0.2$) is shown in Fig. \ref{fig:PhaseDiagram2D_3b}.  
At this point we do not take into account the Zeeman coupling, but consider it
in the subsection below.
The phase boundaries in the absence of an in-plane field (dashed lines) are also plotted,
to make a comparison with the case without the in-plane magnetic field.

Since the in-plane magnetic field breaks the $x$-$y$ symmetry,
the hidden nematic phase denoted by (3) in 
%Fig.~\ref{fig:PhaseDiagram2D_1h}
Fig.~\ref{fig:PhaseDiagram2D_3b} is suppressed under the in-plane field.
Naturally, the second order transition between the isotropic and hidden nematic
phase in the absence of an in-plane magnetic field changes to a first order
transition due to the presence of a symmetry breaking field.
On the other hand, the regions with  parallel nematic order 
are enhanced by the in-plane field, which is also expected.
%their areas in the phase diagram are enlarged by the in-plane field.
%Particulary, the area increased by the field is prominent at higher doping,
%as seen in the phase (2).
While the suppression/enhancement of the hidden/parallel nematic phases
under the in-plane magnetic field are rather robust features,
the orientations of the Fermi surface elongations shown in the insets
depend on the details of the bare band dispersion and the location of the
van Hove singularities.
In the Appendix, we present the DOS at the points (e)-(g) in 
Fig. \ref{fig:PhaseDiagram2D_3b} under the in-plane magnetic field to understand
the relation between the nematic orientations and the van Hove singularities.

\begin{figure*}[tt]
  \includegraphics*[width=0.9\linewidth, clip]{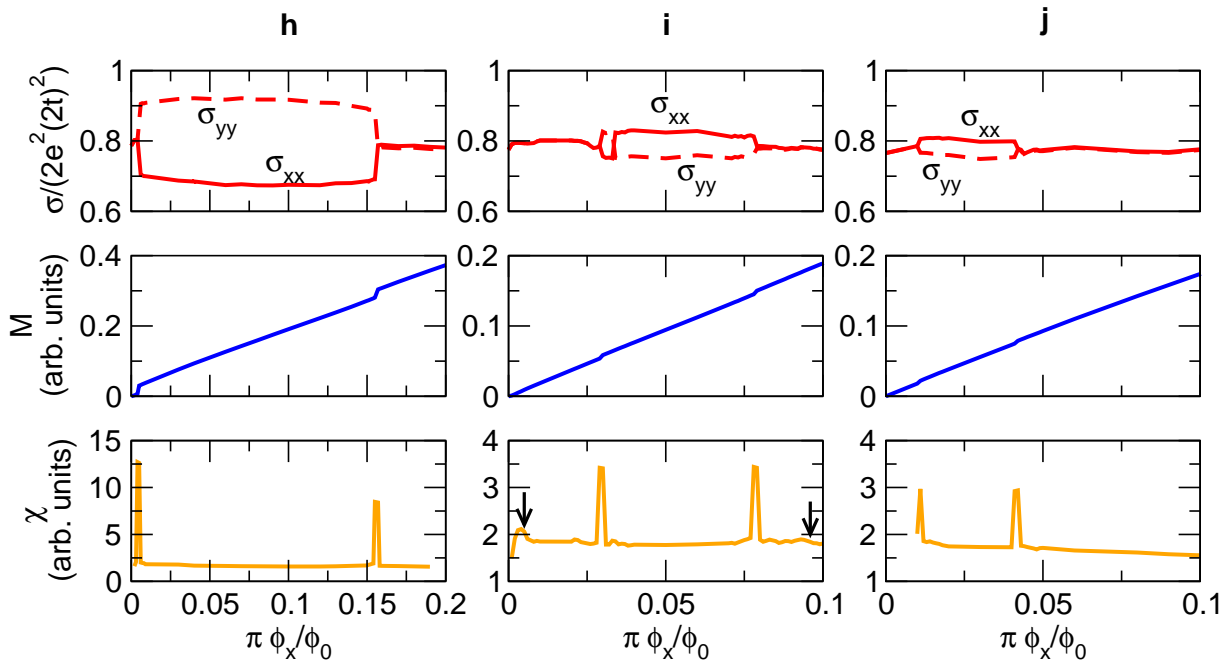} 
  \caption{(Color online) Longitudinal conductivities, 
    magnetization and susceptibility
    as a function of magnetic field strength for  
    $F_{2}/2t = 0.8$, $G_{2}/F_{2} = 0.1$ $\gamma/2t = 1.0$.
    The three panels correspond to the points (h)-(j) as 
    displayed in Fig. \ref{fig:PhaseDiagram2D_1h}.
    %(OPs for panel i are displayed in figure \ref{fig:OP}, 
    The arrows in the susceptibility of the middle panel indicate the perpendicular-isotropic transitions.
    \label{fig:ConMagSusc_hConMagSusc_iConMagSusc_j}}
\end{figure*}

\subsection{Conductivity and Magnetization}
In this section, we present conductivity, magnetization and susceptibility as a function of  
the magnetic field strength $B_x$ (${\bf B} =B_x {\hat x}$).
We then propose a possible connection
between the nematic phase and the phenomena reported in the bilayer ruthenate, Sr$_3$Ru$_2$O$_7$.
%For this comparison, we include the Zeeman coupling to reflect 
%the field driven meta-magnetic transition.
%
In addition to the momentum mismatch due to the magnetic field discussed above,
the Zeeman coupling is important to take into account, since
it acts as a spin dependent chemical potential.
%
%, the phase transition driven by the magnetic field can be 
%regarded as the phase transition driven by the chemical potential but affects 
%two different spins in an opposite way.
%whose direction is denoted as arrows for each spin
%%in Fig.~\ref{fig:PhaseDiagram2D_3b}.
%In addition to the chemical potential shift,
%the effect of tilted magnetic field introduces the in-plane magnetic field
%%%which modifies the phase diagram as shown
%in Fig.~\ref{fig:PhaseDiagram2D_3b}.
%We compute the longitudinal conductivities, magnetization, and susceptibility
%to show the effects of nematicity in the presence of an in-plane magnetic field.
%Here, we include the Zeeman coupling, which leads to an unbalance between up and down
%spin densities.

The longitudinal conductivities are computed using the following standard Boltzmann equation
\begin{equation}
  \sigma_{ii} =  2e^{2} \sum_{{\bf k}, \sigma, \nu=\pm 1 }
  \Big( - \frac{\partial n_{\text{F}}}{\partial E_{{\bf k}, \sigma}^{(\nu)}} \Big)
  \big[{\bf v}^{(\nu)}_{\sigma})_{i}^{2} \tau_{{\bf k}} \big],
\end{equation} 
where we set $\tau_{\bf k}$, originating from impurity scattering, to be constant,
while $n_{\text{F}}$ is the Fermi-Dirac distribution function and 
${\bf v}^{(\nu)}_{\sigma}$ stands for the Fermi velocity.
The magnetization and susceptibility are given by
\begin{equation}
  M = \frac{g \, \mu_{B}}{2} \, \sum_{{\bf k}, \nu}
  \big[ n_{\text{F}}(E^{(\nu)}_{{\bf k}, \sigma = +1}) - n_{\text{F}}(E^{(\nu)}_{{\bf k}, \sigma = -1}) \big],
\end{equation} 
and 
\begin{equation}
\chi = \frac{\partial M}{\partial h}|_{h \rightarrow 0}.
\end{equation}

The conductivity, magnetization, and susceptibility as a function of in-plane
magnetic field 
are shown in Fig.  \ref{fig:ConMagSusc_hConMagSusc_iConMagSusc_j} for 
$F_2/2t=0.8$ and $G_{2}/F_{2}=0.1$.  
The points marked as (h), (i), and (j) in  Fig.
\ref{fig:PhaseDiagram2D_1h} show the values of the chemical potential
and the bilayer coupling used for each panel from left to right, respectively.
%The arrows coming out of the points (h)-(j) indicate that 
%the effect of the magnetic field is the same as shifting
%the chemical potential for up-spin and down-spin in opposite directions
%via Zeeman coupling.
%Since the perpendicular nematic phase is unique signature of our bilayer model,
%we check the signal of perpendicular nematic phase
%in the physical observables.
%Unfortunately, 

Let us first consider the case when the system is in the hidden nematic phase in the absence 
of a magnetic field 
[for instance, close to point (i) in Fig. \ref{fig:PhaseDiagram2D_1h}]. 
As the magnetic field increases, 
the system will undergo several transitions.
The isotropic-hidden nematic phase transition at a weak field
is hardly visible in the conductivity and magnetization
in Fig.~\ref{fig:ConMagSusc_hConMagSusc_iConMagSusc_j} (i),
although the emergence of the nematic order parameter is accompanied by an 
anomaly in the susceptibility as indicated by the left arrow in the susceptibility figure.
At larger magnetic fields isotropic-parallel nematic transitions occur and are clearly signaled by 
the onset and offset of the conductivity anisotropy and the jumps in the magnetization.

On the other hand, consider that the system is in the isotropic phase
in the absence of a magnetic field,
but close to the nematic instability
[see e.g. points (h) and (j) in Fig. \ref{fig:PhaseDiagram2D_1h}].
As the magnetic field is turned on, 
parallel nematic order develops in either the up or the down spin species,
while the Fermi surface of the other spin species stays isotropic.
Then the longitudinal conductivity and the magnetization
show clear indications of the isotropic-parallel nematic transition.
While the preferred direction of the Fermi surface elongation 
in the presence of the symmetry breaking field depends 
on the details of the band structure and the chemical potential, a difference
between $\sigma_{xx}$ and $\sigma_{yy}$
is clearly visible in (h) and (j) 
in Fig.~\ref{fig:ConMagSusc_hConMagSusc_iConMagSusc_j}.
Considering the in-plane field along the $x$ direction,
we find that the conductivity perpendicular to the field ($\sigma_{yy}$)
is higher near half-filling as shown
in Fig.~\ref{fig:ConMagSusc_hConMagSusc_iConMagSusc_j} (h).
Since the Hall conductivity $\sigma_{xy}$ is much smaller than the longitudinal components,
the resistivity is almost inversely proportional to the conductivity. Thus the resistivity 
parallel to the field direction is higher than that perpendicular to the field direction in panel (h), 
which is consistent with the observed anisotropy in Sr$_3$Ru$_2$O$_7$.\cite{Borzi07Science}

\section{Discussion and Summary}

Motivated by the recent experiment on the bilayer ruthenate\cite{Borzi07Science}, 
we address the following question: can one 
explain the magnetoresistive anisotropy within 
the nematic order scenario which successfully describes
both the metamagnetic transitions and the large residual resistivity
\cite{KeeHY05prb,DohH07prl}?
To understand the motivation of our study, one needs to recognize that
the formation of nematic order in a single layer is 
insensitive to the direction of the magnetic field, and thus at  first glance, 
the nematic theory cannot account for all the existing phenomena.
However, here we show that the recently discovered magnetoresistive anisotropy can be explained
within the picture of nematic order when the bilayer coupling is taken into account.
It is essential to note that the in-plane magnetic field leads 
to a relative momentum mismatch between the layers
through bilayer coupling. 
When the field is along one of the in-plane crystal axes,
it breaks the $x$-$y$ symmetry.  
Therefore, the degeneracy of two different nematic orientations
is no longer present, and the system recovers a pure nematic phase with
anisotropic resistivities.

To study the effects of an in-plane magnetic field on the nematic phases in a bilayer system,
we first identify distinct nematic phases in the bilayer system.
While the nematic phase always breaks the $x$-$y$ 
symmetry in the single layer system,
we find that there is  another route to form a different nematic phase called
the hidden nematic phase where the $x$-$y$ symmetry is preserved.
The hidden nematic and the isotropic phase are separated by 
a second order phase transition. While the $x$-$y$ anisotropy is 
absent in both phases, 
 the relative rotational symmetry between the layers 
is broken in the hidden nematic phase.
%.  Since there are two Fermi surfaces in the bilayer system, we call
%the p. 
%the parallel nematic phase.
%(1) The parallel nematic phase, where both Fermi surfaces are distorted along one of crystal axes.
%(2) The perpendicular nematic phase or hidden nematic phase, where the Fermi surfaces preserve the
%4-fold symmetry of the underlying lattice. However, this phase breaks the symmetry between the layers ($\Delta^{\lambda}
%=-\Delta^{(-\lambda)} \neq 0$, where $\lambda$ represents the layer index).

The effect of the in-plane field is rather straightforward when one recognizes
the importance of the bilayer structure in Sr$_3$Ru$_2$O$_7$.
As discussed above,
since the in-plane field is an $x$-$y$ symmetry breaking field, 
there is no spontaneously broken
$x$-$y$ symmetry in the presence of an 
in-plane magnetic field. It is clear that the two-fold
degeneracy is no longer present, and 
domains cannot be formed under the same mechanism 
as described in Ref. \onlinecite{DohH07prl}:
one of the Fermi surface elongations is 
energetically preferred over the other for any small amount
of in-plane field.
Thus the system recovers the intrinsic anisotropy, 
unless it is in the hidden nematic phase.
%Comparing the phenomena observed in the bilayer ruthenates and show that
Based on the analysis of longitudinal conductivities 
and magnetic susceptibilities, we propose
that Sr$_2$RuO$_7$ is  close to
the parallel nematic instability in the absence of a magnetic field.

One may question the validity of the mean field theory adopted in the current paper, since it is widely
accepted that Sr$_3$Ru$_2$O$_7$ is a strongly correlated material with a putative
quantum critical point.
It is true that a mean field theory breaks down close to a quantum critical point
due to large fluctuations, and it is plausible that large fluctuations are important to determine
an effective Hamiltonian. However, one should note that we attempt to describe the ordered
state and its first order transition to the isotropic phase.
Since a mean field theory works quite well deep inside
an ordered state due to negligible fluctuations, 
and the effects of fluctuation on a first order transition
are not as important as for a  second order transition,  
we argue that the mean field theory with nematic order qualitatively captures the 
phenomena discussed above.

There are strong indications that a magnetic field tuned
nematic phase bounded by isotropic regions exists in the bilayer ruthenate. 
In addition to the phenomena discussed above, a recent scanning tunneling microscopy experiment
under a $c$ axis 
field revealed the splitting of singularities in the local DOS across the 
metamagnetic transition.\cite{takagi07prl}
However,  further experiments to detect a direct Fermi surface anisotropy, 
such as scanning tunneling microscopy under in-plane magnetic fields,
are highly desirable.\cite{DohH07prb} 
A microscopic mechanism for
the formation of a nematic phase beyond the effective 
model Hamiltonian\cite{Kee08},
 and the effects of disorder
in relation to a putative quantum critical point are also important subjects 
for theoretical studies which we will address in the future.

\begin{acknowledgments}
We thank Stephen Julian and Eduardo Fradkin for useful discussions.
This work was supported by NSERC of Canada, Canada Research
Chair, Canadian Institute for Advanced Research, and 
Alfred P.~Sloan Foundation (HYK).
\end{acknowledgments}

\section{appendix}

\begin{figure*}[tt]
  \includegraphics*[angle=0, width=0.9\linewidth, clip]{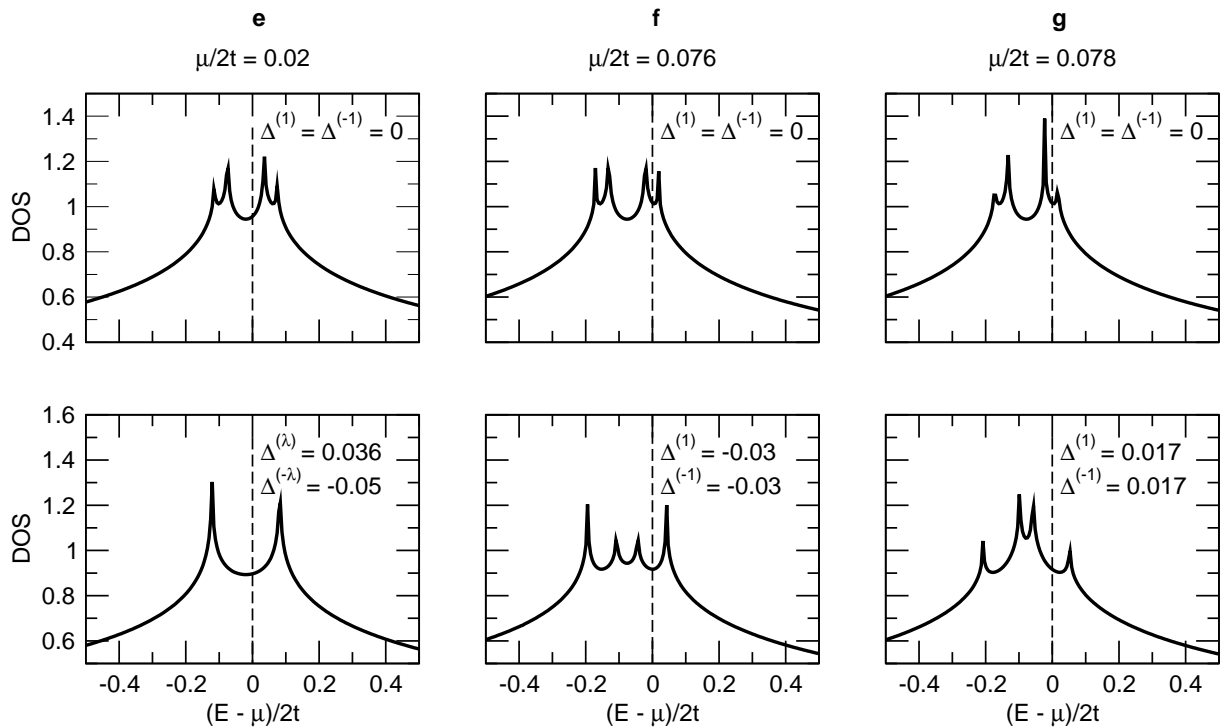} 
  \caption{DOS in the presence of a magnetic field,
    $\pi \, \phi_{x} / \phi_{0} = 0.2$, for various points along the 
    line $t_{\perp} / 2t = 0.075$ in Fig. \ref{fig:PhaseDiagram2D_3b} (DOS in units of 2t/E). 
    We set $F_{2} /2t = 0.8$ and  $G_{2}/F_{2} = 0.1$.
    The first row shows the DOS of the bilayer tight-binding model, where the nematicity is forced to be zero
    to see the effects of finite nematicity on the DOS, which is shown in the bottom row.
    \label{fig:DOSAndFS_3a}}
\end{figure*}

%bb=1.8cm 4.2cm 24.7cm 17.8cm, 
Here we present the DOS (Fig. \ref{fig:DOSAndFS_3a}) in the presence of 
an in-plane magnetic field, to understand the correlation between the preferred
direction of Fermi surface elongation and the locations of the van Hove singularities.
We set $\pi \, \phi_{x} / \phi_{0} = 0.2$,  $t_{\perp} / 2t = 0.075$,  $F_{2} /2t = 0.8$, 
and  $G_{2}/F_{2} = 0.1$. The values of chemical potential used for the DOS correspond to those at 
the points (e)-(g) in Fig. \ref{fig:PhaseDiagram2D_3b}. 
In the absence of an in-plane magnetic field, there are two peaks separated by
the bilayer coupling, $2 t_{\perp}$. However, the in-plane magnetic field splits
each of the peaks into two peaks such that there are four singularities in the DOS.
The top row shows the DOS of the underlying tight-binding 
bilayer model in the presence of an in-plane magnetic field,
where the magnetic field induces a splitting of the tight-binding singularities into asymmetric peaks.
Here, we force the nematicity to be zero in order to see the effect of finite nematicity on the DOS,
which is shown in the bottom row of Fig. \ref{fig:DOSAndFS_3a}.
The three bottom panels represent three different ways of avoiding 
van Hove singularities in the presence of a finite magnetic field, 
which can be found by comparing the top and bottom panels for each case. 
In (e), a finite nematic order with two different orientations for each layer
turns four peaks into two peaks such 
that the Fermi level is further away from the modified  
singularity.  In (f) and (g), 
parallel nematic order not only splits the two peaks near the Fermi level
further apart from each other, but also shifts the weight of the DOS between the singularities.
In all three cases, nematic order leads to a further separation between the Fermi level
and  the nearby van Hove singularity.
This analysis helps us understand the orientations of the Fermi surface elongation for the
particular
set of parameters used here, and one should bear in mind that the preferred direction of the Fermi surface
distortion is sensitive to the parameters used in the nematic theory, since van Hove singularities depend on
the details of the band structure.


\begin{thebibliography}{13}
\bibitem[{\citenamefont{footnote}(2007{\natexlab{b}})\citenamefont{
  footnote}}]{footnote}
\bibinfo{note}{Here we do not take into account  possible diagonal nematic phase discussed
in Ref. \onlinecite{DohH06prb} and \onlinecite{Kao05prb}}.
\end{thebibliography}
\end{document}